# The use of incentives to promote Technical Debt management

Terese Besker
Computer Science and Engineering, Software Engineering
Chalmers University of Technology
Göteborg, Sweden
Besker@chalmers.se

Antonio Martini
University of Oslo
Programming and Software Engineering
Oslo, Norway
antonima@ifi.uio.no

Jan Bosch
Computer Science and Engineering, Software Engineering
Chalmers University of Technology
Göteborg, Sweden
Jan.Bosch@chalmers.se

## ABSTRACT

***Context***: When developing software, it is vitally important to keep the level of technical debt down since it is well established from several studies that technical debt can, e.g., lower the development productivity, decrease the developers' morale, and compromise the overall quality of the software. However, even if researchers and practitioners working in today's software development industry are quite familiar with the concept of technical debt and its related negative consequences, there has been no empirical research focusing specifically on how software managers actively communicate and manage the need to keep the level of technical debt as low as possible.

***Objective:*** This study aims to understand how software companies give incentives to manage Technical Debt. This is done by exploring how companies encourage and reward practitioners for actively keeping the level of technical debt down and whether the companies use any *forcing* or *penalizing* initiatives when managing technical debt.

***Method:*** In a first step, this paper reports the results of both an online survey provided quantitative data from 258 participants and interviews with 32 software practitioners. In a second step, this study set out to specifically provide a detailed assessment of additional and in-depth analysis of Technical Debt management strategies based on an encouraging mindset and attitude from both managers and technical roles to understand *how*, *when and by whom* such strategy is adopted in practice.

***Results:*** Our findings show that having a Technical Debt management strategy (specially based on encouragement) can significantly impact the amount of Technical Debt in the software.

***Conclusion:*** The result indicates that there is considerable unfulfilled potential to influence how software practitioners can further limit and reduce Technical Debt by adopting a strategy based explicitly on an encouraging mindset from managers where they also specifically dedicate time and resources for Technical Debt remediation activities.

## 1. INTRODUCTION

When developing software, it is vitally important to keep the level of technical debt (TD) down since it is well established from several previous studies that TD can, for example, lower the development productivity [1], decrease the developers' morale [2], and compromise the overall software quality [3] and even lead to a crisis point when a huge, costly refactoring or a replacement of the whole software needs to be undertaken [4].

The TD metaphor was first introduced by Ward Cunningham [5] to illustrate the need to recognize the potential long-term negative effects of immature code that is sub-optimally implemented during the software development lifecycle. This debt must be repaid with interest in the long term [6].

Even if the concept of TD and its negative consequences are quite well known to software engineering (SE) practitioners today, there is always a risk that TD remediation tasks gets down-prioritized or neglected by the practitioners since today's' software practitioners face increased pressure from management to reduce the development time and thereby to reduce the costs of the development [7]. On the other hand, it is, at the same time, important to deliver high-quality software with as little TD as possible. This balancing act between implementing and delivering the software as fast as possible and at the same time spending time and effort on both avoiding introducing TD in the first place and conducting TD refactoring activities of already implemented software gets particularly demanding.

Like other professionals, software engineers' work outcomes, attitudes, and work behaviors are influenced by their company's corporate culture and the managers' mindset. This means that managers can have an outsized impact on the overall software development process by adopting different management strategies and using techniques for controlling and directing software engineers to achieve predetermined goals.

In recent years, the use of different strategies in behavioral interventions has become more prevalent [8]. In a literature review, this study initially identifies four different strategies that managers can adopt to impact how practitioners work with TD. Besides *encouraging* employees by, for example, introducing training programs that focus on raising awareness and enhancing knowledge about specific desired behavior, there are also other strategies managers can implement to impact [9] and motivate [10]-[11] their employees In general, one mechanism managers use to impact practitioners' work is an incentive program, where a specific behavior is recognized and *rewarded* [12]-[13]. Oppositely, managers can also use disincentive programs to *penalize* an undesired or destructive behavior [14]. Further, managers can similarly implement explicitly *forcing* requirements and rules, where all concerned employees must fulfill and adapt to in order to continue their work or, for example, to deploy their implementations and continue developing new tasks [15].

However, to the best of our knowledge, this is the first study to investigate empirically how common and important different management strategies are when specifically managing TD in today's software development industry.

This study is carried out in two steps. First, we study four different TD management incentive strategies to manage TD (addressing RQ 1-4). Secondly, based on the findings in the first step, we provide a detailed assessment of additional and in-depth analysis of one of these strategies (encouragement) in order to understand *how*, *when and by whom* such strategy is adopted in practice (addressing RQ 5-7). In particular, this study examines the following seven research questions:

**RQ1:** How common is an *encouraging* attitude to keep the level of TD down, and do software engineering practitioners perceive this TD management strategy as an effective or desirable strategy?

**RQ2:** How common are *rewarding* incentives to keep the level of TD down, and do software engineering practitioners perceive this TD management strategy as an effective or desirable strategy?

**RQ3**: How common is it to use a *forcing* mechanism to keep the level of TD down, and do software engineering practitioners perceive this TD management strategy as an effective or desirable strategy?

**RQ4**: How common are *penalizing* disincentives to keep the level of TD down, and do software engineering practitioners perceive this TD management strategy as an effective or desirable strategy?

**RQ5:** What specific TD management activities are encouraged and who encourages these activities?

**RQ6:** In what situations or under what circumstances are practitioners encouraged to address TD?

**RQ7:** How are practitioners encouraged to address TD?

This paper reports the results of two sets of surveys where the first survey is an online survey providing quantitative data from 258 respondents, and the second survey provides data from 72 respondents. The result is also based on qualitative data from interviews with initially 32 software practitioners from seven software companies, followed by yet another round of four interviews.

To the best of our knowledge, no known empirical research has focused on exploring the relationships between TD and different management strategies, and the contribution of this subject in this paper is fivefold: First, we show how commonly used each of the investigated strategies are within today's software industry. Second, we present a TD management quadrant model that presents four different TD management strategies and illustrates its strategies and tactics, together with recommendations on how to implement them. Third, our result shows that a TD management strategy can significantly impact the amount of TD in the software. Fourth, when surveying how commonly used different TD management strategies are, we found that only the encouraging strategy is, to some extent, adopted in today's' software industry. Lastly, our result clearly shows that there is a misalignment in *how* and *when* the managers perceive they encourage the development teams to address TD, in comparison to how and in what situations or under what circumstances the teams perceive being encouraged by their managers. Taken together, these findings provide valuable insights into the role management has on the way practitioners address TD during their software development work.

This manuscript was originally and partly published at the Third International Conference on Technical Debt, held jointly with ICSE [16]. The delta of this manuscript over the prior published paper is based on an additional empirical extension of the previous study, where this manuscript includes an in-depth value-added analysis of the results derived in the first study in order to provide a more detailed and comprehensive understanding and perception of the first sets of results. This extended study also addresses the comments we received from the anonymous reviewers during the first submission.
This manuscript has been extended to include three additional research questions (RQ5-7), where the results of these questions are derived from a totally new set of independent data collection.
The related Research section has been extended to be broader and more carefully cover additional related research publications to address the additional research question.
The Methodology section is updated to also include the additional step of the study, together with an illustration and description of how the different steps of this study relate to each other.
Further, in both the Result and the Discussion sections of this extended version of the study, several new findings and results are added and discussed. These additional results highlight the previous results and further strengthen the understandability of the first set of results, thereby bringing a finer granularity to our understanding of the practice.

The rest of this paper is organized as follows. Section 2 presents the used conceptual framework, and Section 3 introduces related work. Section 4 describes the research methods in detail. Section 5 presents the research results. Section 6 discusses the findings, and Section 7 presents threats to the study's validity, and Section 8 concludes the study.

## 2. CONCEPTUAL FRAMEWORK

As briefly described in Section 1, organizations can use different management strategies to influence employees' working behaviors. Initially, we search for research publications addressing different strategies that managers can use for influencing practitioners' working behaviors and attitudes. Based on this review's outcome, we have depicted a conceptual model presented in Fig.1 and formulated the research questions based on this framework. The framework illustrates four main different management strategies; *Encouraging*, b) *Rewarding*, c) *Forcing*, and d) *Penalizing*, where each of them connects to one of the research questions presented in Section 1. The conceptual framework with its four main strategies is proposed under the rationale that these four strategies are potentially important and can influence TD's reduction and potentially are being used by managers to manage the amount of TD during the software development work.

Since encouragement was found as the most important activity used in practice by practitioners (see section 4), this strategy encompasses a more detailed perspective, where further aspects are covered related to *how*, *when*, and *by whom* encouragement is carried out in practice.

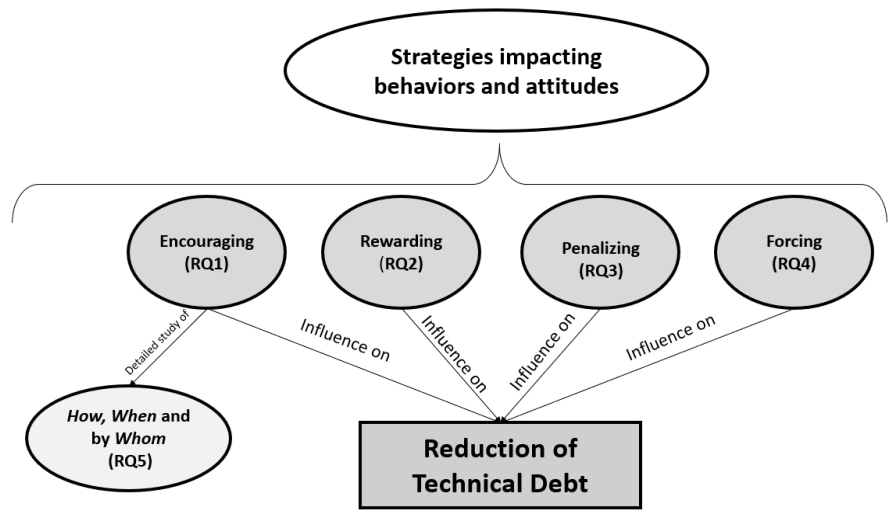

**Figure 1:** Conceptual framework

## 3. BACKGROUND AND RELATED WORK

This section presents related work concerning incentive and disincentive programs in today's software engineering field, followed by the different management strategies, as illustrated in the conceptual framework in Fig. 1.

### 3.1 Incentives and disincentive programs in SE

An incentive program addresses a planned activity designed to motivate employees (individuals or teams) to achieve specified and predetermined organizational goals or objectives within a specific time frame. At the same time, a disincentive program is the antonym of the incentive program and discourages employees (individuals or teams) from performing specific activities [17].

Commonly, software development projects are measured using financial indicators. Typically, the incentive program aims to give bonuses to managers who run their projects with a high-profit margin or within the budget timeframe [18]. There is no research to date on how other software engineering roles, such as, for example, developers, testers, and architects, are included in incentive or disincentive programs in general and, more specifically, TD management.

### 3.2   Encouraging activities

Encouraging employees is an important part of being a leader where the leader highlights and compliments specific desired actions and where the leader also provides constructive criticism if needed. Managers' behaviors provide an important message to employees, meaning, for example, that a high level of creativity and innovation result from managerial behaviors [19] where the relationship between employees and their managers has a significant bearing on employees' work-related attitudes and behaviors [9]. By default, just encouraging employees does not include any direct rewards.

### 3.3   Rewarding incentive

Several studies show that reward and recognition programs can positively influence motivation, performance, and interest within an organization [12],[13]. The overall goal with reward and/or recognition programs is to foster teamwork, boost employee loyalty, and ultimately facilitate the development of a wanted culture that rewards a specific behavior [12]. The practitioners (individuals or teams) who fulfill the goals get a predefined reward. A reward program can, for example, recognize developers who adopt suggested techniques, and thereby the reward incentive gives a significant boost to those who deploy best practices, where the achievement, for example, can be rewarded by a badge [20] or by a gift card or a monetary reward.

### 3.4   Forcing mechanisms

The strategy based on forcing mechanisms refers to mandatory rules and requirements that need to be fulfilled and followed by the practitioners to demonstrate adherence to methodologies, rules, regulations, guidelines, or best practices [15]. This could be exemplified by a situation in which mandatory rules and requirements are not met. Hence, it forces the practitioners to go back and alter the software before being allowed to continue with, for example, adding additional features or deploying the software. Examples of commonly adopted rules and requirements from an SE perspective are not allowing any bugs in the software and requiring that the software be thoroughly tested before deployment and the code fully reviewed and that it follows the coding standards.

### 3.5   Penalizing disincentive

Organizational penalty or punishment) is a pervasive phenomenon in many companies and organizations [21] that yields penalties for an undesired behavior and is also a disincentive strategy. Penalization refers to when managers apply a negative consequence or the removal of a positive consequence following an employee's undesirable behavior, intending to decrease the frequency of that behavior [14]. According to Wang and Zhang [21], some software development organizations have adopted punishment measures in an attempt to improve software developers' performance, reduce software defects, and hence ensure software quality. Their result shows that while a penalty mechanism helps to reduce software defects in daily coding activity, it fails to achieve programmers' maximum work potential. In their study [21], penalty rules were introduced when software developers were tracked submitting unsuccessful submissions, which caused monetary fines for the individual developer.

Since there is a current gap in research addressing how different TD management strategies impact how practitioners work with TD, this study built on research conducted in other domains and examines the current state of different management strategies in the SE field. Our work is, therefore, different from the studies mentioned above in several aspects: We a) provide results derived from data from a real software development environment rather than discussion without empirical evidence as support, b) we combine both qualitative and quantitative methods, c) our study investigates four different management strategies, and d) our investigation primarily focuses on TD.

### 3.6    TD Management activities

TD management facilitates decision-making about the need to remove or avoid a TD item and the most appropriate time to do this [22]. There are several different TD management activities, which may significantly impact the amount of TD within a software system. Since TD has a significant negative impact on the software development work from several different perspectives, our previous research [23] show that it is essential to actively prevent introducing TD in the software in the first place and to iteratively and continuously conduct TD tasks when the TD already is introduced in the software [24]. This means that targeting and encouraging software practitioners to perform such activities (e.g., avoid and remove TD) may significantly impact reducing the harmful effects of TD.

However, to remediate or refactor TD, the identified TD items first need to be tracked and prioritized, preferably using an official backlog. Our previous research [7] indicate that when practitioners use so-called "shadow backlogs," the TD items potentially may result in being overlooked during the prioritization process, with the result that the TD items will remain in the software. Moreover, there are situations where deliberately taking on TD may be a strategically sound move since such conscious decisions sometimes may, e.g., increase the ability to cut development time, and thereby enable fast feedback from customers and increase revenue [25], [26], [27]. Therefore, and on top of investigating strategies to keep the level of TD down, this study also addresses the extent to which management does exactly the opposite by assessing if the teams are encouraged to take on additional TD deliberately.

Besides encouraging the avoidance and the remediation of TD, it is also essential to address *when* TD refactoring activities should occur and whether the practitioners are empowered to make such decisions on their own or if such decisions must be taken together with managers [7]. The above-presented activities and situations are used in the survey in step 2 of the study.

### 4.   METHODOLOGY

As visualized in Fig. 2, this study used a combination of quantitative and qualitative research approaches. The research design was divided into two main steps, including a total of 11 different phases. As illustrated in the figure, the first six phases were conducted in step 1, which refers to this publication's original study [16]. The following phases (7 to 11) were conducted as an extension of that study.

The research approach used in step 1 is characterized by an exploratory approach where several different incentive strategies are studied. This step's outcome is used as input to the following step 2, which is characterized by a conclusive study approach, where this step primarily focuses on one type of incentive program from step 1.

The following sections describe each step with its phase together with its related research methods.

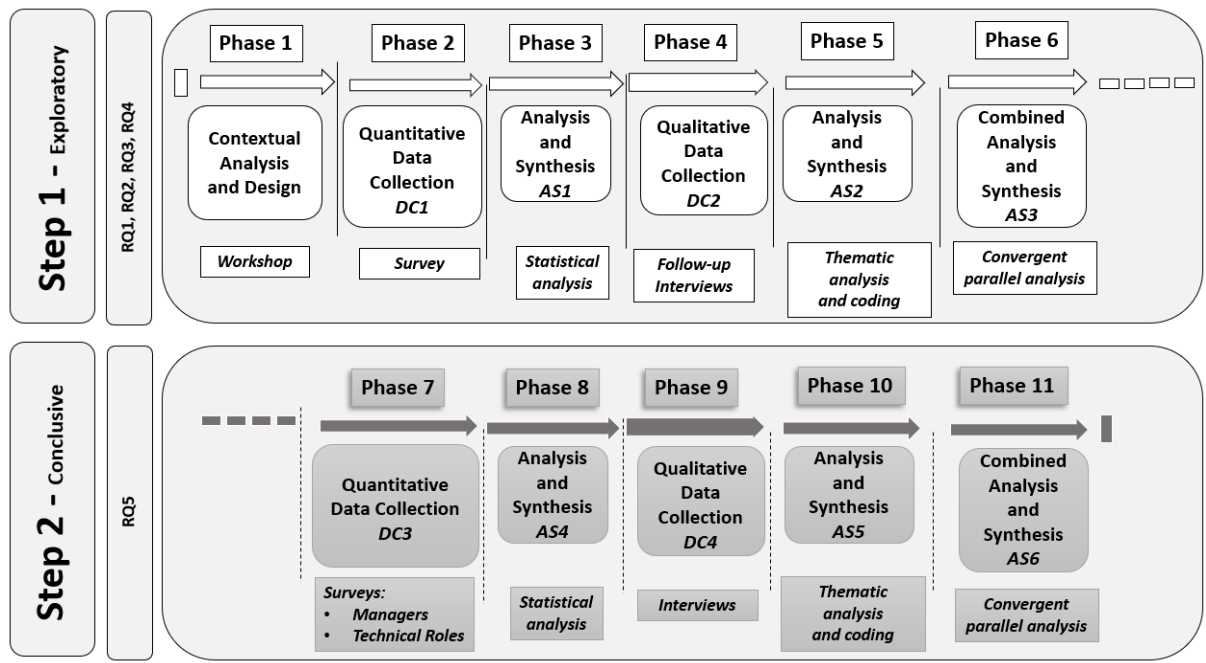

**Figure 2:** Research Design

## 4.1 Step 1 – The exploratory part of the study

The first step of this study's exploratory nature aims to answer RQ1-RQ4 and is described in the following six sub-sections.

### 4.1.1 Phase 1 – Contextual Analysis and Design

The study was first presented and discussed during a workshop with software practitioners from seven software companies within our industrial network. All companies had an extensive range of software development. This phase's outcome is the research model shown in Fig. 1, which directed the design, data collection, and analysis of the following phases.

This step of the study adopted a selection of respondents using mainly a purposive sampling technique [28] of software professionals. The aim of primarily using a purposive sampling technique was to select relevant and suitable candidates for the study. Taken together, out of the 258 respondents in the survey, 21 respondents came from LinkedIn invitations from software engineering groups, and the remaining 247 respondents came from our network of seven of our industrial software partners. All those seven companies had an outspoken strategy and goal of enhancing their TD remediation work and thereby strived to reduce the negative effects of TD.

### 4.1.2 Phase 2 – Quantitative Data Collection (DC1)

The data collected in this phase was supported by an online web survey designed and hosted by SurveyMonkey. The motivation for using a survey in this part of the study was to reach a high level of generalizability based on a large population of software professionals [29]. According to the guidance provided by Czaja and Blair [30], the first draft of the survey was tested by four industrial practitioners (developer, manager, project owner, and software architect) and by two Ph.D. candidates in order to evaluate the understanding of the questions and the usage of common terms and expressions [30]. During this evaluation, we also monitored the time needed to complete the survey.

The survey invitations were emailed directly to seven companies within our networks, all located in Scandinavia, having an extensive range of software development, and invitations were also published on software engineering-related networks on LinkedIn. The surveys were anonymous, and participation in the surveys was voluntary. Across all these collaborators, 312 respondents began the survey, and 258 respondents answered all questions. Due to high completion rates (~83%), we decided to reject the incomplete responses, according to the guidelines proposed by Kitchenham and Pfleeger [31].

The first part of the survey gathered descriptive statistics to summarize the respondents' backgrounds and their companies. These data are reported in detail in Appendix A.

The survey included respondents having different roles, where 49% were developers-/programmers/software engineers, while 25% were software architects. Approximately 78% of the software architects and 59% of the developers/programmers/software engineers had more than ten years of experience. The most common size of the software development team was 6–10 members (36%), and most (32%) systems were, on average, 5–10 years old from their initial design.

The second part of the survey included the four survey statements (ST) presented in Table 1 to facilitate quantitative answering the RQs presented in Section 1 (when fully answering the RQs, we used quantitative data from this phase combined with qualitative data as described in section 4.4).

For each of the statements, the respondents were asked to indicate their level of agreement on the 6-point Likert scale; Strongly Agree, Agree, Somewhat Agree, Somewhat Disagree, Disagree, and Strongly Disagree.

**Table 1: Characteristics of the sample survey – All Roles in Step 1**

| ID | Statement | Addressing RQ |
|---|---|---|
| **ST1** | Our team or I am explicitly rewarded if TD is kept down. | 2 |
| **ST2** | Our team or I am explicitly penalized if TD is not kept down. | 4 |
| **ST3** | Our team or I am explicitly forced to keep the level of TD down (i.e., to be allowed for deployment) | 3 |
| **ST4** | Our team or I am explicitly encouraged if TD is kept down. | 1 |

### 4.1.3 Phase 3 – Analysis and Synthesis (AS1)

The survey data were analyzed quantitatively, that is, by interpreting the numbers obtained from the answers. The data were analyzed using descriptive statistics and graphically visualized using diverging stacked bar charts.

### 4.1.4 Phase 4 – Qualitative Data Collection (DC2)

In this stage, the second round of data was collected, where 32 software practitioners were focus-group-interviewed. As suggested by Runeson and Höst [32], this study employed the technique of semi-structured interviews, where the questions were planned but not necessarily asked in the same order as they were listed. These interviews were used to obtain detailed information about the interviewees' perceptions and interpretations of the study topics.

All interviews were focus-group interviews based on guidelines by Krueger and Casey [33], stating that this method is specifically suitable, serving as a source of follow-up data to assist a prior used data collection method: "*The researchers need the information to help shed light on quantitative data already collected*."

In total, we interviewed seven companies, where each interview included between four to seven interviewees. Altogether, we interviewed 32 experienced software development professionals with roles as architects, developers, product owners, and managers. All interviewees had participated in the previous survey. For confidentiality, interviewees and their companies are kept anonymous.

All interviewees were asked for recording permission before starting, and they all agreed to be recorded and to be anonymously quoted for this paper. Each interview lasted between 105 and 120 minutes and was digitally recorded and transcribed verbatim. Examples of interview questions for each RQ are presented in Appendix A.

Before the interviews started, the previous survey's compiled results were presented to the respondents (using graphical illustrations such as bar diagrams and graphs). This presentation allowed the respondents to relate the interview questions more easily to the results of the survey.

The interview questions were designed to a) increase the understanding of the survey results, b) ensure that the survey questions were understood and interpreted as intended and uniformly, c) confirm the survey results, and d) understand the survey results' implications. The questions were developed to cover the same taxonomies as the previous survey to validate the findings of the survey.

### 4.1.5   Phase 5 – Analysis and Synthesis (AS2)

This stage focused on analyzing the data collected in the previous phase. The analysis and the synthesis of the data were analyzed using thematic analysis [34].

First, the transcriptions from the recorded interviews were manually coded using established guidelines in the literature [35]. The tracking of the links between the codes and the quotations was supported by a Qualitative Data Analysis tool called Atlas.ti.

Secondly, based on the research taxonomy, a coding scheme containing four broad categories and 19 individual codes was developed. For example, the citation "*I think it sounds like the reward would be the best way of keeping the measure of technical debt down*" was coded as "Rewarding."

To ensure that the coding was performed consistently and reliably, two authors of this study synchronized the codings' output, as suggested by Campbell et al. [36]. The outcome of this analysis process (where the mapping between the different hierarchical categories and individual codes is presented graphically) is attached in Appendix A.

### 4.1.6   Phase 6 – Combined Analysis and Synthesis (AS3)

In the sixth phase of the study, we combined the results we received in the previous quantitative and qualitative data collection and analysis phases.

## 4.2   Step 2 – The conclusive part of the study

The second step of this study's conclusive nature aims to answer RQ5, RQ6, and RQ7. The background of the research conducted in this step is based on the results derived in Step 1, where about 60% of the survey respondents state directly being encouraged by managers to keep the level of TD down (see section 5.1.1).

As a result of this finding, in step 2, we, therefore, further specifically investigate a TD management strategy based on explicit encouragement to understand better how and by whom this encouragement is carried out in practice.

In this step of the study, we conducted a single, case-study, following the guidelines provided by Runeson and Höst [32]. We consider the case as embedded as multiple units of analysis were studied within the case (different groups of roles). The selected case company is a large technological development supplier with around 385.000 employees operating worldwide. The participating respondents work at several different sites, operate in different countries, and have different managers. This case company was selected due to its ongoing initiative addressing TD during their software development work, making it suitable for studying an incentive strategy based on encouragement. For confidentiality reasons, the company name has been anonymized in this study.

### 4.2.1   Phase 7 – Quantitative Data Collection (DC3)

The data in this phase was collected using a similar approach as in phase 2, Step1, where the quantitative data were collected using the online service provided by Surveymonkey. However, this phase collected quantitative data using two totally new different sets of surveys; one for managers and one for technical roles.

The first parts of each of the surveys gathered descriptive statistics to summarize the backgrounds of the respondents. These data presented in detail in Appendix A. Altogether, 26 managers and 46 technical roles participated in the surveys.

In these two sets of surveys, all respondents were asked to indicate their agreement level on a 4-point Likert scale (Strongly Agree, Agree, Disagree, and Strongly Disagree). The assessed statements in both of the surveys were of similar character but were phrased in slightly different ways. The questions' goal was to address the same topics but represented by both a manager and a technical role perspective.

The surveys included other questions on other topics related to TD management, such as process, tools, etc., which are not relevant for this study.

As illustrated in table 2, we used the same statements but altered the phrasing depending on whether the questions were asked to managers or technical roles.

Table 2: Characteristics of the sample survey – All Roles in Step 2

| *Manager Roles phasing:* | *Technical Roles phrasing:* | *Statements* |
|---|---|---|
| **I encourage the software development teams to:** | **My manager encourages my team to:** | - Avoid and Remove TD<br>- Assess and report TD in the official backlogs to prioritize and remove it<br>- Deliberate taking on TD if they get benefits out of it (e.g., to speed up delivery) |
| | **My team colleagues encourage me to:** | - Avoid and Remove TD<br>- Assess and report TD in the official backlogs to prioritize and remove it<br>- Deliberate taking on TD if they get benefits out of it (e.g., to speed up delivery) |
| **When is my team encouraged to remove TD?** | **When is my team encouraged to remove TD?** | - Whenever they/we want<br>- When they/we have extra time, budget, or human resources to be allocated<br>- When they/we have a specific amount of time dedicated to TD removal (e.g., 10, 20 %, etc.)<br>- When they/we provide a business case for removing TD (e.g., reporting on costs, risks, and benefits of removing or keeping TD) |

### 4.2.2 Phase 8 – Analysis and Synthesis (AS4)

This phase's data was collected using a similar approach as in phase 3, Step1, where the survey data were analyzed quantitatively. In this phase, the managers' data and the technical roles from the two sets of surveys were first analyzed separately and, after that, meta-analyzed together.

### 4.2.3 Phase 9 – Qualitative Data Collection (DC4)

For this phase's data collection, four different semi-structured interviews were conducted, using the same settings and approaches, as described in section 4.1.4.

We interviewed two interviewees with technical roles (developers), followed by three additional interviews with four managers (Project owner, Chief Product Owner, Process Manager, and Head of Architects). All interviewees had previously taken the survey in part 2, and the results from the surveys were presented to the interviewees during the interviews.

### 4.2.4 Phase 10 – Analysis and Synthesis (AS5)

This stage focused on analyzing the data collected in phase 9, using the same approach as in phase 5, Step1.

### 4.2.5 Phase 11 – Combined Analysis and Synthesis (AS6)

This phase combined the previous quantitative and qualitative data collection results, using the same approach as in phase 6, Step1.

# 5. RESULTS AND FINDINGS

This section presents the results of this study, where the first four sub-sections (5.1-5.4) present the result for RQ1-4, followed by sections 5.5 and 5.6, which addresses RQ 5-7.

In the survey used in Step 1, the participants were asked to rate their agreement with four statements (ST1-ST4) using a 6-point Likert Scale. The ratings provided by the respondents for each of the survey statements are presented in Fig. 3 and further described in the following subsections 5.1 to 5.4.

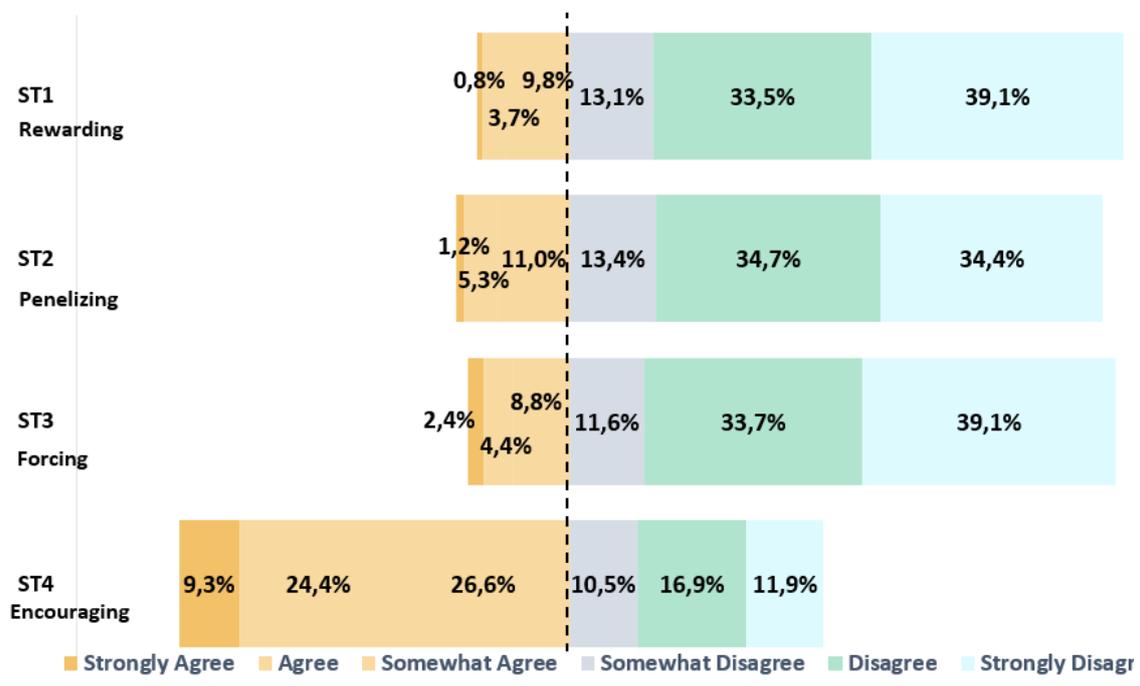

**Figure 3:** Summary of the responses to the survey statements in Step 1.

## 5.1 Encouraging strategy (RQ1)

The first research question investigates how common an *encouraging strategy* is to keep the level of TD down and how software engineering practitioners perceive this TD management strategy.

### 5.1.1 Survey results
When looking at the results from the different statements in Fig. 3, it is evident that one of the statements excels compared to the other statements.
What stands out in the figure is the statement assessing whether the respondents are encouraged to keep the level of TD down (ST4), where 60.3% (9,3% strongly agree, plus 24,4% agree, plus 26,6% somewhat agee) of the respondents agree to some extent of being encouraged, and 39.7% disagree to some extent of being encouraged to keep the level of TD down.
A remarkable result of this statement is that 11.9% of the respondents strongly disagree with being encouraged to keep the level of TD down.

### 5.1.2 Effective or desirable strategies
The practitioners' attitudes toward introducing TD or conducting TD remediation tasks were described as being guided and targeted by the mindset and the attitudes from the management side where recognition of leaders and peers was important. This meant that when management focused its attention on TD's importance, the employee was encouraged to focus his or her work in the same direction.

All the interviewees considered an encouraging managing strategy of TD as highly effective and impactful. Several of them described that this strategy clearly could be more emphasized within their organizations and thereby also have a more significant impact on the amount of TD in their software.

### 5.1.3 Tactics for encouragement

Several actions were identified as examples of encouraging activities from management to keep the level of TD down.

Several of the interviewed companies strived to continuously raise awareness about the concept of TD and its related negative consequences as an encouragement to keep the level of TD down. Some companies ran satisfyingly dedicated educational sessions to explicitly address how to avoid TD's introduction in the first place. One interviewed software architect said, "*We* [the architectural team] *try to help some teams here that we are close to, and we try to teach them how to write good APIs.*" Yet another interviewee from another company echoed this view, "*I think of education. I think a lot of debt is introduced because of a lack of knowledge.*"

The managers for some teams in one of the interviewed companies had set aside a specific amount of working time within each sprint to allow explicitly (without imposing on the developers) spending time on TD remediation activities together with other software-improving activities. This dedicated time slot encouraged the involved engineers (such as testers, developers, and architects) to focus on TD issues in every sprint as an incorporated part of their overall working process.

## 5.2 Rewarding Incentives (RQ2)

The second research question addresses how common rewarding incentives are to keep the level of TD down and how software engineering practitioners perceive this TD management strategy.

### 5.2.1 Survey results

As illustrated in the first statement (ST1) in Fig. 3, only 14.3% of the respondents agree to some extent (0,8% strongly agree, 3,7% plus agree, plus 9,8% somewhat agree) with being explicitly rewarded when keeping the level of TD down. Thus 87.7% of the respondents state that they are not explicitly rewarded for it. What stands out in this data is that only two (2) respondents state that they strongly agree with being rewarded when keeping the level of TD down; meanwhile, a hundred (100) respondents state that they strongly disagree with being explicitly rewarded if they keep the level of TD down.

### 5.2.2 Effective or desirable strategies

None of the interviewed companies had an incentive program where employees were rewarded for any (not only TD specifically) explicit behavior. The interviewees' thoughts differed as to whether adopting rewarding incentives is an effective or desirable strategy to keep the level of TD down.

Some interviewees were skeptical of explicitly highlighting and rewarding specific working activities and behaviors such as TD remediations since they thought keeping the level of TD down should be an activity that comes with the craftsmanship of software development and the working pride of software engineers. For example, one interviewee said, "*It sends the wrong signals if you start to reward or penalize individuals or teams for something that should be a normal part of their daily work. Because I see that paying off technical debt should be done as a part of your daily work to the extent that is possible*".

On the other side, several other interviewees argued that a TD managing strategy based on rewards could be effective since rewards can motivate practitioners to manage TD further. One such interviewee said, "*I guess it is a good way. If you have this kind of reward now and then, you keep up the awareness that this is important. It is something of a signal from the organization.*"

### 5.2.3 Tactics for rewards

Concerns regarding the different appropriate types of rewards were widespread. Some proposed (since none of them had any incentive programs in place) rewards such as monetary compensation, extra holidays, and pizzas to the teams. Meanwhile, other interviewees said a reward does not have to be tangible; it could be a simple acknowledgment since official praise, and thus an enhanced reputation, is considered equally important: "*A reward does not have to be money. You get some reputation that will actually be enough reward in itself.*"

Nevertheless, even if a rewarding incentive has the best intention to decrease the amount of TD in the software, such initiatives could easily be misused by causing a counterproductive backlash where, for example, practitioners primarily focusing on TD remediation tasks in order to get the rewards or focusing only on the TD items that are easy to refactor, and thereby focusing less on other tasks and goals, which can harm the overall implementation or delivery of the software. As one interviewee put it, "*How do you make sure that you don't end up having a person that's just fixing technical debt [in order to get the reward].*"

Yet another concern that was expressed by several interviewees refers to the possibility of manipulating such reward systems by first introducing a large amount of TD and, after that, refactor it in order to get the reward. "*Well, the problem is that it's very easy to game it. I could introduce a large technical debt item and then refactor it [to get the reward].*" Further, an incentive program rewarding the reduction of TD after the TD has first been identified, suppose that in order to get the reward, the reduction of TD must initially be identified and tracked. This suggests that even if the TD is recognized before the manager has identified it, the practitioners could postpone refactoring the TD to get the reward. As one interviewee said, "*If you see the debt going up, and you fix it directly, you don't get any reward. I mean, it's easy to cheat.*"

Taken together, if an incentive program for TD remediation should be introduced, such a program must be carefully designed to avoid counterproductive results that instead generate even more TD; and it is important to design it in an impartial way that is not easy to manipulate.

### 5.3 Forcing mechanisms (RQ3)

The third research question aims to assess the forcing mechanism to keep the level of TD down and how software engineering practitioners perceive this TD management strategy.

#### 5.3.1 Survey results

When assessing whether the respondents are being forced to keep the level of TD down, the result from the second statement (ST3) in Fig. 3 shows that 15.6% of the respondents agree to some extent (2,4% strongly agree, 4,4% plus agree, plus 8,8% somewhat agee) with this statement and that 84.4% of the respondents disagree to some extent with being forced to keep the level of TD down.

#### 5.3.2 Effective or desirable strategies

None of the interviewed companies had any forcing rules or requirements related to TD. Notably, all the companies had other forcing rules related to their software development processes, such as following code standards, documentation requirements, and performing tests. These rules applied primarily to specified mandatory activities and requirements that had to be fulfilled for the delivery to be viewed as complete and further activities to take place.

Even if the TD and its negative effects are known to the software engineers, it can be challenging to get the time and budget from the management side for refactoring the software. Here, a positive side to a forcing TD management strategy was described in terms of empowerment. Such a strategy would give the practitioners authority to conduct mandatory TD remediation tasks without arguing and motivating the action to the managers. For example, one interviewee said, "*I think the motivation is very high to take care of all technical debt, but we don't have the funding to do it. It's no problem motivating the software engineers to fix it, they really want to do it, but we don't have the funding most of the time. I would like the forcing part stating that you actually need to fix this.*" Yet another finding was that the forcing of TD remediation activities seems to become more vital for companies adopting shared ownership of their software product portfolio, where several teams collaborate on the same software without having strict ownership of the components. Even if the different teams had guidelines that, to some extent, addressed the management of TD, there was no uniformity among the teams regarding how they managed TD since the used guidelines were based only on recommendations without any direct obligations or forcing mechanisms. For example, one interviewee said, *"Since we have shared ownership of the code, then we need to have a certain set of common rules to follow to have a good look and feel of the code independent of which team that was in there last. And these rules need to be forced by someone."*

No direct negative effects of having forcing rules or requirement related TD were identified in the study. However, several concerns about how such TD managing forcing rules should be designed and implemented was raised by the software practitioners. For example, it is vitally important to define which rules, measurements, and responsibilities should be applied when adopting a forcing strategy. Together, they address how these goals can be achieved. For example, one interviewee said, *"You need good rules that people understand, and they need to understand what they shall do."* This view was echoed by another interviewee who stated, *"But without measurements, you don't have anything to go on. And if everybody is aware of they are being measured, then they will not mess up"*.

### 5.3.3 Tactics for implementing a forcing strategy

Another view on the enforcement of TD activities was described as a transition from an encouraging strategy to a forcing strategy.

Several interviewees recommended that a company first focus on encouraging initiatives, and if such a strategy were conceived as not enough, this forcing strategy could be implemented. To directly implement a forcing strategy was not recommended by the interviewed companies. One company described their history going from an encouraging strategy to adopting a forcing strategy (however, this was not forcing of TD management): *"It depends on the scale on the organization. A couple of years ago, we didn't force that much. We were encouraged, and that was because we had some sort of thought of ownership, and I would say product pride in our product, and that was our sort of encouragement. It was your baby, and you wanted to be proud of it, and that's driving bits."*

## 5.4 Penalizing Disincentives (RQ4)

The fourth research question set out to investigate how common *penalizing* disincentives are to keep the level of TD down and how software engineering practitioners perceive this TD management strategy.

### 5.4.1 Survey results

Looking at the third statement (ST2) in Fig. 3, it is apparent that 17.5% of the respondents agree to some extent (1,2% strongly agree, 5,3% plus agree, plus 11% somewhat agree) with being explicitly penalized when not keeping the level of TD down, and thus 82.5% of the respondents state that they are not penalized if they do not.

### 5.4.2 Effective or desirable strategies

Even if the survey showed that respondents are being penalized, none of the interviewees in the study were familiar with any penalizing activities within their companies, causing, for example, monetary fines and salary reductions. All of the interviewees had direct negative attitudes toward implementing a TD management strategy based on penalizing practitioners or teams who fail in keeping the level of TD down. Commenting on penalization strategies, one of the interviewees said, *"No, I don't like, for example, lowered salary because we don't meet the expectations."*

### 5.4.3 Tactics for penalizations

Several different concerns were raised about a penalizing strategy for managing TD. Initially, to penalize an undesired behavior where, for example, TD is introduced (deliberate or un deliberate) or when identified TD are not refactored in a wanted way or within a specified period, there must be clear rules for the software practitioners to follow. These rules can potentially be the same type of rules as the rules formulated in the forcing strategy, but they can also be of a different character.

However, there are several different issues to consider if a penalizing strategy should be implemented to facilitate the management of TD. Several interviewees raised the importance of establishing fair, adequate, and succinct rules that should be clearly conveyed, understood, and followed by all the concerned employees.

Initially, there must be a general understanding and definition of TD among all concerned employees. An interviewee addressing this issue said, *"For penalizing, then people are going to sweep it under the carpet and say, "Oh, I don't know about any technical debt'."*

Further, there also needs to be someone to be accountable for defining the rules and someone who should be responsible for tracking, identifying, and penalizing the individual employee or the team. The justice of the rules must also be addressed appropriately, meaning that there must be clear rules on "whom" to penalize. One example to illustrate this can be when a project manager orders a developer to introduce TD intentionally due to time restrictions. The penalizing rules must be clear on whom to penalize in such a situation: the person who orders the introduction of TD or, for example, the developer who actually implements the suboptimal solution. One of the interviewed developers highlighted such a scenario by describing that TD sometimes has to be intentionally introduced because of a requirement coming from outside the one's team: *"The penalizing part, it's often just like it's not in your control if you want to keep the debt down fully, and some of this is probably your fault, but often it's something from outside your team that's saying: 'No, you have to do this now* [introduce the TD delibertely].*'"*

## 5.5 The perception of Encouragement - Survey results

This section reports the findings from the second step of the study, based on the first step's findings. One of the key findings in the first step was that about 60% of the survey respondents stated that they were encouraged by managers to keep the level of TD down (see section 5.1.1). This section addresses the research questions RQ5, RQ6, and RQ7, which all provide a more detailed assessment and an additional and in-depth analysis of the TD management strategy based on encouragement.

First, we report the findings related to the perception of being or providing encouragement, together with the findings related to the encouragement of different professional roles, followed by a subsection addressing the different situations *when* the refactoring of TD is encouraged. The final subsection presents the findings related to *how* TD management's encouragement is carried out in practice.

Since encouragement is a human activity based on communication that involves both a "sender" and a "receiver," wherein the sender conveys a message, and the receiver gets that message, we have asked both managers and technical roles to state to what extent they perceive that they send the message (managers) and to what extent they receive the message (technical roles).

### 5.5.1 The perception of receiving or providing encouragement (RQ5)

This sub-section addresses the fifth research question to understand the extent to which different specific TD management activities are encouraged and *who* encourage these activities.

This section addresses encouragement of addressing TD, seen from three different role perspectives:
a) how managers perceive providing encouragement to technical roles, b) how technical roles perceive receiving encouraged by managers, and c) how colleagues having technical roles encourage each other within the teams to address TD.

Further, to provide a higher granularity of the encouragement of TD management activities, we collected data using the three different TD encouraging activities: 1) avoid and remove TD, 2) assess and report TD in the official backlogs to prioritize and remove it, and 3) deliberately taking on TD if they get benefits out of it (e.g., to speed up delivery).

The respondents' quantitative summary statistics for each of the survey statements are presented in Fig. 4 and further reported and described together with the qualitative results, from the interviews, in the following sections.

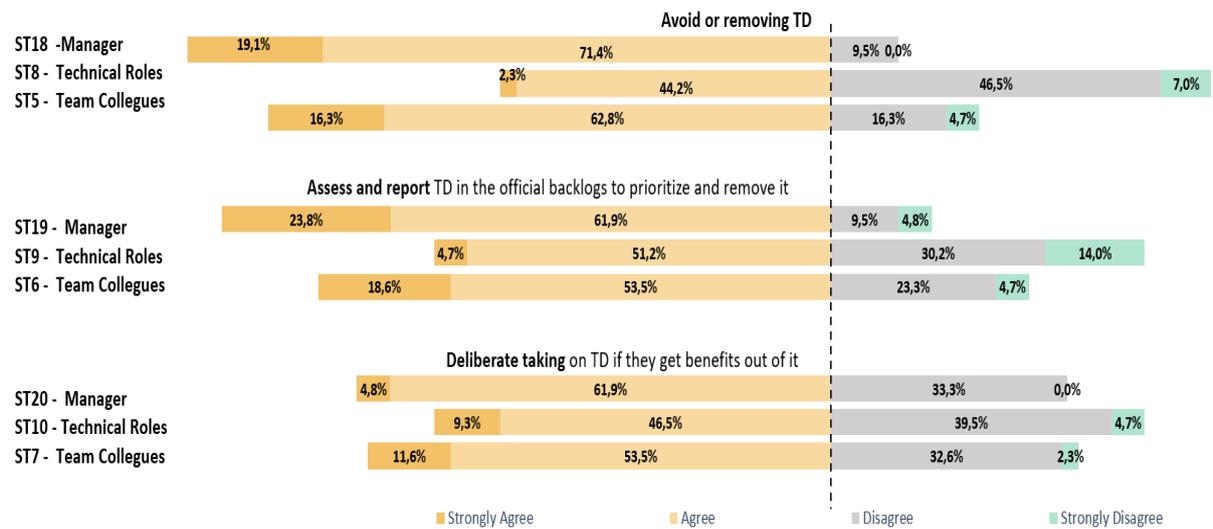

**Figure 4:** Summary of the responses to the survey statements reflecting the perception of receiving or providing encouragement

*Encouraged to avoid and to remove TD:* The first three upper stack bars in Fig. 4 (ST18, ST8, and ST5) address the encouragement to *avoid or removing TD*, and by looking at these bars, it is apparent that the results illustrating this activity differ significantly between the different groups of respondents.

As illustrated in the first stack bar (ST18), about 91% (19,1 % strongly agree plus 71,4 % agree) of the managers who took the survey agree to encourage the software development teams to avoid or remove TD. In comparison to that, "only" 46 % (2,3% strongly agree plus 44,2% agree) of the technical roles (in bar ST8) perceive that they are being encouraged by their managers. However, about 79% (16,3% strongly agree plus 62,8% agree) of the technical roles (in bar ST5) perceive that they are encouraged by their team colleagues to avoid or to remove TD.

This is a rather remarkable result indicating that the managers perceive that they encourage the technical roles' teams to avoid or remove TD. However, the technical roles do not seem to receive this encouragement to the same extent. Though, it is apparent from the results that the technical roles perceive being encouraged by their team colleagues to avoid and remove TD.

*Assess and report TD in the official backlogs to prioritize and remove it:* As illustrated in Fig. 4, bar ST19, about 86% (23,8% strongly agree plus 61,9% agree) of the managers stated in the survey that they encourage the teams to assess and report TD in the official backlogs to prioritize and remove it.

When surveying the technical roles to what extent they perceive receiving encouragement from managers to perform this activity, about 56 % (4,7 % strongly agree plus 51,2 % agree) of the technical roles (in bar ST9) perceive that they are being encouraged by their managers, where also 14 % of the respondents strongly disagree to this statement.

Further, by looking at the results in bar ST6, it is apparent that about 72 % (18,6 % strongly agree plus 53,5 % agree) of the technical roles perceive that they are encouraged by their team colleagues to avoid or to remove TD Assess and report TD in the official backlogs.

*Deliberate taking on TD if they get benefits out of it:* As shown in Fig. 4, bar ST20, about 67% (4,8% strongly agree plus 61,9% agree) of the managers stated in the survey that they encourage the teams to deliberate taking on TD if they get benefits out of it. This encouragement from the managers was received by 56 % of the technical roles (in bar ST10), and 65 % of the technical roles also perceive being encouraged of this activity by their team colleagues (bar ST7).

When summarizing and analyzing the three different activities together, it becomes clear that managers perceive they encourage the technical roles to *avoid and remove TD* to the greatest extent. However, this result is thought quite contrary to how the technical roles perceive this activity since this activity is actually the activity that the technical roles perceive as being the least encourage one from the manager side among the three different listed activities. However, this activity seems to be the most common activity that team colleagues encourage each other to address.

Further, the activity that managers most strongly agree to was that they encourage the technical roles to assess and report TD in the official backlogs (23,8% strongly agree). Meanwhile, seen from the technical role perspective, this activity was the activity the technical roles most strongly disagree with being encouraged to address (14% strongly disagree).

Taken together, these results suggest that there is a quite extensive misalignment between how managers perceive they encourage the technical roles to address TD and how the technical roles perceive being encouraged by their managers. However, it is clear that the technical roles encourage each other to address TD, to a relatively large extent.

### 5.5.2 Individual professions – grouped as Technical roles (RQ5)

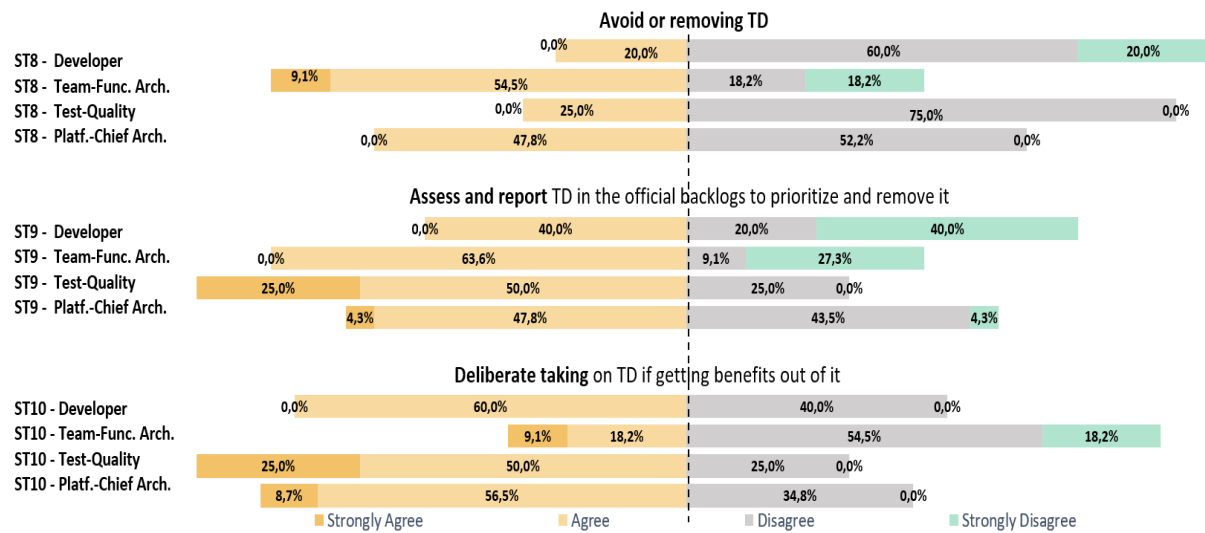

**Figure 5**: **Summary of the individual professions responses to the survey statements reflecting the perception of receiving encouragement**

To further study how the technical roles' teams are encouraged by their managers to address TD, this section presents the survey results from each of the individual roles in the earlier reported results, where they were analyzed as a group. These results are illustrated in Fig. 5.

The distinct roles that we earlier grouped as "technical roles" (in section 5.5.1) are a) developers, b) team- functional architects, c) test and quality engineers, and d) platform-chief architects.
Further, we use the same TD encouragement activity as earlier reported in section 5.5.1.

***Encouraged to avoid and remove TD:*** The first four upper stack bars in Fig. 5 (ST8) visualize the extent to which the different individual roles agree their teams to being encouraged by their managers to avoid or remove TD.

From the data, it is apparent that the different investigated professions perceive being encouraged (from managers) quite differently to avoid or remove TD. What stands out when comparing the result is, for example, that none (0%) of the developers strongly agree to this statement and that only 20 % of them agree to be encouraged by their managers to avoid or to remove TD.

Further, when looking at these results, it is also apparent that Test and quality engineers perceive being limited encouraged by their managers, where 25% agree, and the remaining 75% disagree with being encouraged to avoid or removed TD.

Meanwhile, about 64 % (9,1% strongly agree plus 54,5% agree) of the team-and functional architects agree to the same statement. However, this profession's results are quite spread, where, for example, 18,2 % strongly disagree with this statement.

*Assess and report TD in the official backlogs to prioritize and remove it:* As illustrated in Fig. 5, bars ST9, 0 % of both the developers and the team- and functional architects strongly agree of being encouraged to assess and report TD in the official backlogs and also quite a few of these professions strongly disagree to this statement (40 % contra 27,3%). However, three-thirds (75 %) of the surveyed Test and Quality engineers agree to be encouraged by their managers to assess and report TD in the official backlogs.

*Deliberate taking on TD if they get benefits out of it:* As shown in Fig. 5, bars ST10, 60 % of the developers perceive that they are being encouraged by their managers to deliberate taking on TD if they get benefits from it  (0 % strongly agree plus 60 % agree). However, the results of this activity show that only about 27 % (9,1 strongly agree plus 18,2 agree) of the Team- and functional architects agree with this statement. Also, 18,2 % of them strongly disagree with being encouraged to deliberate taking on TD if they benefit from it.

Taken together, when looking at these results, it is clear that different individual roles perceive the extent of their managers' encouragement differently, where, for instance, developers commonly strongly disagree with all three statements. However, due to a relatively small sample size of each of these professions, caution must be applied, as the findings might not be generalized to a larger sample of professions. Further data collection is required to determine whether this result holds for a larger population of each type of profession.

### 5.5.3   The situation when the teams are encouraged to remove TD (RQ6)

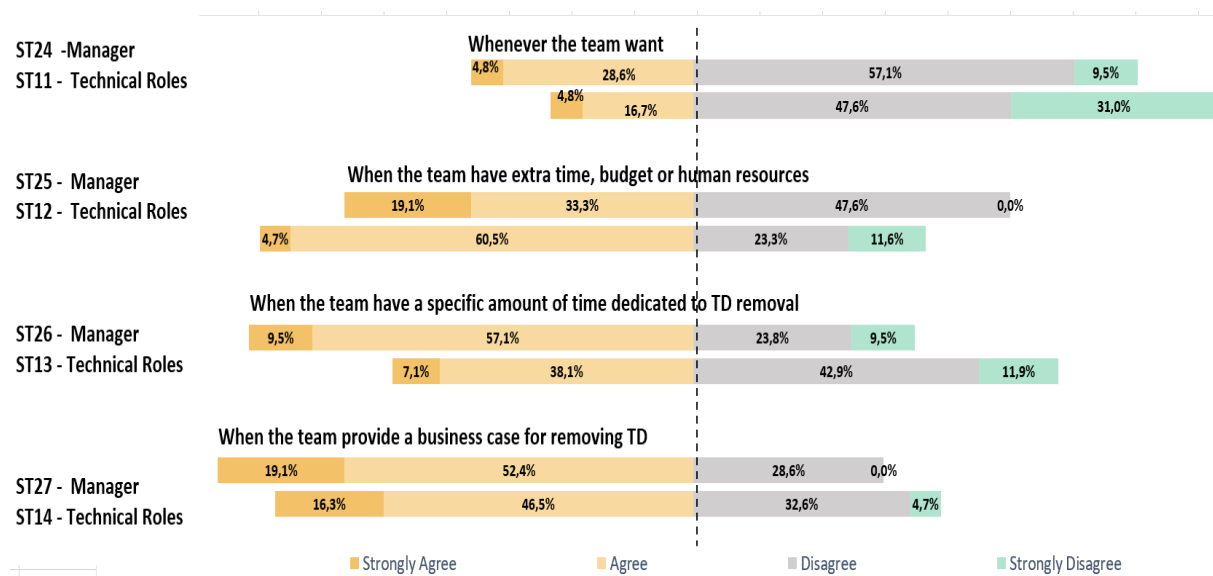

**Figure 6:** Summary of the responses to the survey statements reflecting the situation when the teams are encouraged to remove TD

This section reports the survey results, addressing different situations or under what circumstances when the managers perceive that they encourage the teams to remove TD and receive such encouragement. As illustrated in Fig. 6, we assess four different situations and circumstances where the encouragement of removing TD may take place; a) whenever the team wants, b) when the team has extra time, budget, or human resources, c) when the team has a specific amount of time dedicated to TD removal, and d) when the team provides a business case for removing TD.

***Encouraged to remove TD whenever we/the team want:*** As shown in Fig. 6, bar ST24, about one-third of the managers perceive (4,8% strongly agree plus 28,6% agree) they encourage the teams to remove TD whenever they want. This view is shared by about 21 % of the technical roles (4,76% strongly agree plus 16,67 % agree), as illustrated in ST11. However, one should also notice that a substantial part of the technical roles (about 79 %) disagree with being encouraged by managers to remove TD whenever they want (47,6 % disagree plus 31 % strongly disagree).

***Encouraged to remove TD when we/the team have extra time, budget, or human resources:*** When assessing the result of ST25 and ST12 in Fig. 6, the results show quite interestingly that the technical roles perceive receiving more encouragement to remove TD when they have extra time, budget, or human resources compared to what the managers report that they actually encourage the teams to. This result is the only result among all the assessed statements where the managers perceive that they encourage less than how the technical roles perceive their managers encourage them.

***Encouraged to remove TD when we/the team have a specific amount of time dedicated to TD removal:*** It can be seen from bar ST26 and ST13 in Fig. 6 that about two-thirds of the managers agree that they encouraged the teams to remove TD when they have a specific amount of time dedicated to TD removal (9,5 % strongly agree plus 57,1 % agree). However, only about 45 % of the technical roles report that their managers encouraged them to do so (7,1 % strongly agree plus 38,1 % agree), meaning that about 55 % of the technical roles disagree with this statement.

***Encouraged to remove TD when we/the team provide a business case for removing TD:*** The last two stack bars (ST27 and ST14) in Fig. 6 report the result related to the encouragement of removing TD when the teams provide a business case for removing TD. By looking at these stack bars, it is evident that by comparing this result with the previous comparisons, the perception of these statements seems to differ less between the managers and technical roles. ST27 shows, for instance, that 71% (19,1 % strongly agree plus 52,4 % agree) of the managers agree to encourage this action. Meanwhile, about 63 % of the technical roles perceive this type of encouragement from their managers.

When analyzing the four different activities together, the results indicate that the technical roles perceive that they are more encouraged to remove TD by their managers when they have extra time or resources available or when they can provide a business case for TD refactoring activities. Moreover, the technical roles seem to perceive less encouragement from their managers to remove TD whenever they want. Taken together, when analyzing the statements which relate to when the TD refactoring is encouraged, it is evident by the presented findings that there is a misalignment between the perception between the manager perspective compared to the technical role perspective, related to the different situations and circumstances when conducting TD refactoring activities are encouraged.

## 5.6 The perception of Encouragement: qualitative results and discussion

After the quantitative data was collected and analyzed, the results were presented to a subset of the respondents during four different interviews with both technical roles and managers (separately).

The survey results were described as quite surprising by two of the interviewed managers, where for example, one Chief Product Owner (CPO) said, "*I would not expect that they [the managers] put so much effort on encouraging TD management.* Another interviewed manager offered a potential explanation for the relatively high percentage of the perception of encouragement from the manager's side as "*The managers themselves think they encourage TD management a lot since it is a very high percentage…I think the managers 'want' to have this attitude of encouraging the removal of TD.*"

Moreover, during these follow-up sessions, one of the interviewed managers concluded the overall results as "*The perception of the managers and the technical people is completely different, but the teams show some awareness, so they* [the teams] *deal with it* [TD] *anyway.*"

Contrary, during the interviews with respondents having technical roles, concerns were expressed about the encouragement of addressing TD from their managers' side. Even if they perceived that they received some encouragement from their managers, the TD related tasks were commonly down-prioritized in favor of implementing new features instead, which was perceived by technical roles (especially developers) as a lack of encouragement.

When interviewing a Chief Product Managers (CPM), it was revealed that they often do not agree to prioritize large TD issues (for example, related to architecture refactoring) because of the lack of a good business case to support it, which should be provided by technical roles (especially architects) and because of the lack of a suitable solution proposal to remove the TD. This was reported by a CPM with a technical background, which decreases the likelihood that, as proposed by other technical roles in the interviews, managers would down-prioritize TD refactoring because of lack of technical understanding.

When asking what kind of TD was encouraged to be removed, we found a difference that could explain the divergence of perception between managers and technical roles. Both managers and technical roles revealed that they encouraged and are encouraged respectively to remove TD that hinders the implementation of the features in the immediate future, which in most cases is represented by "small" TD issues, or other issues that can be fixed by dedicating a limited percentage of development time (e.g., 10-20%). However, such time is not enough to remove larger TD, such as the architectural ones. Large architectural refactorings, for which often packages of epic dimension are required and should be prioritized against features at a high management level, are still often down-prioritized. Consequently, managers feel that they encourage to remove (small) TD, while technical roles feel that they are not encouraged to remove (large) TD. This difference in perceiving encouragement might have caused the misalignment reported in the data.

Another issue brought up by POs during the interviews was that TD needs to be prioritized against interests from several stakeholders in complex projects. The projects' complexity often leads to wrongly estimated time to implement the features: then, the extra time needed for the features ends up, in practice, decreasing any time reserved to refactor TD. This means that, even if a percentage of time is reserved to remove TD, it is not actually enough, and if the pressure from implementing features is too high. Inferring that, TD should be prioritized at a higher management level and when the projects' budget is decided. Our results further reveal that TD is not, in fact, often taken into consideration as a variable when project resources are allocated.

Taken together, the misalignment of encouragement between the managers and the technical roles may be because managers, in general, perceive that they encourage the technical roles, and even if the technical roles receive this encouragement, they are not provided with additional time and resources to actually address TD in reality.

However, with limited interviewees, caution must be applied, as the findings might not be transferable to all different technical roles and all manager roles.

## 6. DISCUSSION AND LIMITATIONS

Since no research to date was found in the SE research field on how different TD management strategies can be applied to keep TD down, this research provides novel results.

In the first step of the study, four main strategies were identified in initial literature research, which was particularly interesting when portraying how software managers can influence and impact the software engineers' attitudes and working behaviors with TD.

As illustrated in Fig. 7, we propose a model describing the different identified and investigated TD management strategies. The model spans four quadrants and is named "The Four TD Management Quadrants," which outlines that TD managing strategies can be either an incentive/disincentive character focusing on either a desired or undesired behavior.

This model offers an exploration of different TD management strategies, with their different strategies and tactics, and it also describes both how the strategies relate to each other and how they differ. This model can assist managers in deciding which strategy to adopt and support transition plans, shifting from one strategy to another.

Among the different studied strategies, the result shows that today's software companies most commonly use a TD management strategy based on the encouragement of employees, where 60% of the respondents in the survey state that they are, to some extent, encouraged to keep the level of TD down. Meanwhile, the other investigated strategies, such as using a strategy based on forcing mechanisms or adopting incentive or disincentive programs, were rarely used by the companies.

Further, among the investigated companies, there was a strong belief that both the encouraging and the reward TD management strategy would be valuable to decrease further the amount of TD in the software; meanwhile, the forcing and penalizing strategies were not considered as desired and constructive strategies.

One motivating finding is that practitioners conceive that the attitudes and mindset toward TD remediation tasks from their managers significantly impact the way they address TD. Further, and perhaps the most striking results in this study are that a TD managing strategy based on encouragement has a significant impact on the way practitioners work with TD.

Though, since the result shows that still quite a lot of the respondents to some extent do not agree with being encouraged (40%) to focus their effort on TD remediation tasks, this result shows that there is an unfulfilled potential for managers to impact how practitioners can reduce TD by adopting a TD management strategy based on encouragement and without having to introduce forcing mechanisms or strategies based on rewards or penalizations.

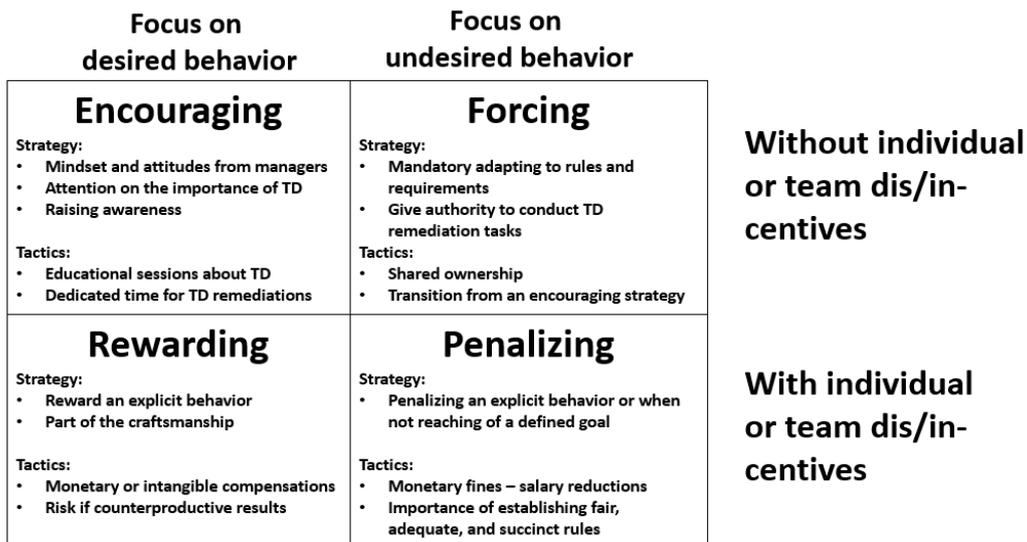

**Figure 7:** The Four TD Management Quadrants Model

The second step of the study primarily focused on the output from the first step to provide more in-depth information about *how, when, and by whom* TD management is encouraged.

It is evident from this step of the study that there are significant differences in how managers perceive that they encourage the technical roles compared to how the technical roles perceive being encouraged by the managers. Commonly, the managers perceive that they encourage addressing TD significantly more often compared to how the technical roles perceive receiving this encouragement from their managers. The result also indicates a misalignment in when or under which circumstances TD refactoring activities should be carried out.

Several different factors could explain the misalignment of TD management's encouragement between the technical roles and their managers. In particular, we found that the higher-level managers encourage teams indeed to remove short-term and smaller TD issues. However, they do not often prioritize large TD issues to be refactored. This happens because of a lack of business cases and viable refactoring solutions. In a sense, this can also be seen as the *managers not feeling encouraged enough to prioritize TD refactoring* at a higher level. In turn, the down-prioritization causes the technical roles (especially developers) to also feel not encouraged to remove TD due to lack of dedicated time.
These relationships are illustrated in Fig. 8, where the current successful (green arrows) and unsuccessful (red arrows) encouragement practices across the different roles are illustrated.

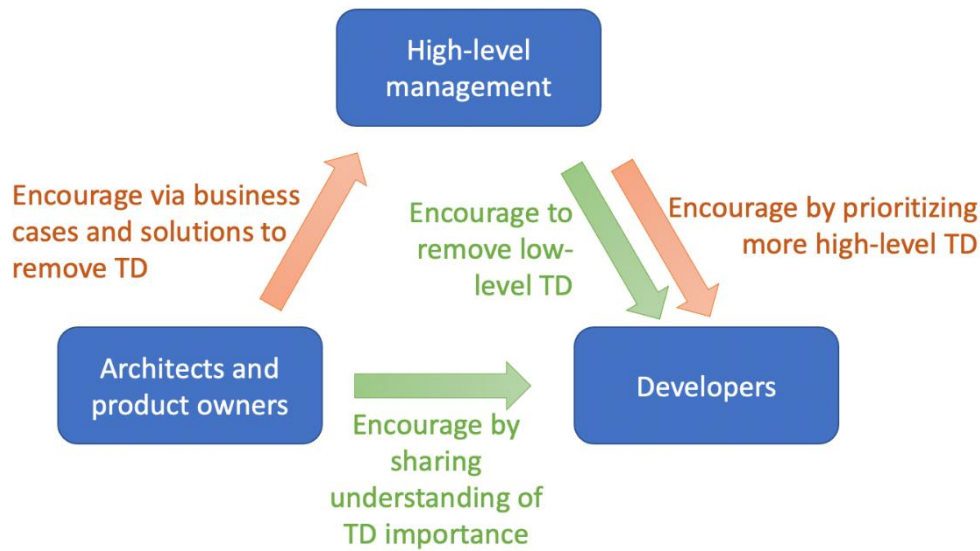

**Figure 8:** Encouragement strategies between roles

Although architects and POs agree that TD is important to be fixed, they report struggling to provide the right business motivation for TD to be prioritized (for example, the quantification of the TD interest). In addition, feature pressure and the complexity of the projects and stakeholders tend to cause unplanned work that then eats up also the time dedicated to remove TD.

Given the reported results, we foresee four possible encouragement practices, which need to be further investigated:
1) Technical Debt should be recognized as a variable in project planning in order to better protect time for the eventual and unavoidable occurrence of TD.
2) Architects and POs need to motivate higher-level management to prioritize the removal of large TD items. To do so, more emphasis can be put on creating business cases and quantifying the TD interest.
3) High-level managers should prioritize more larger TD issues against the features to convey the message to the developers that TD is indeed an important entity to be taken into consideration.
4) We also found that companies are often unaware of the extent of misalignment between employees' perception and top management. Once this gap was showed to the participants, they were keener to revise their encouragement strategy

In conclusion, the success of encouragement across the roles seems to be closely related to how the TD is communicated and prioritized.

## 7. THREATS TO VALIDITY

Several vital threats to validity necessitate a cautious interpretation of the results of the present study. We have chosen a classification scheme to distinguish between different aspects of validity and threats to validity provided by Runesson and Höst [32]. This scheme includes four aspects of validity: construct validity, internal validity, external validity, and reliability.

Construct validity reflects the extent to which the studied operational measures represent what the researchers have in mind and what is investigated according to the stated research questions [37]. This is commonly one of the main threats to validity in surveys, as the respondents might interpret the survey questions and other terms differently. To mitigate this threat, we provided the respondents with the following description of TD before answering the questions: "*Technical Debt is a metaphor that describes a real-life phenomenon, and it provides a way of talking and reasoning about difficulties related to software development and software maintenance. Below is a brief description of what Technical Debt is: Technical Debt (TD) is usually described as the non-optimal code or other artifacts related to software development that gives a short-term benefit but causes a long-term extra cost during the software life-cycle.*" Further, this study could possibly suffer from internal validity by affecting our ability to explain the phenomena that we observed [21] accurately. However, to mitigate this threat, we triangulated both survey's findings by conducting follow-up interviews validating the derived results. Additionally, to minimize the threat of misunderstanding the different topics in the survey, we initially conducted two pilot studies (one for each survey) with industrial practitioners. Understanding the terms was also addressed during the follow-up interviews. External validity focuses on the extent to which it is possible to generalize the findings. There is always a risk in surveys that the sample is biased, and for this topic, a potential threat refers to the demographic and cultural distribution of response samples. As reported in Section 4.1.2, we mainly investigated companies from the Scandinavian area in step 1, which may have a cultural impact on respondents' experiences and views on penalizing, rewarding, forcing, and encouraging management actions. The results could, therefore, potentially be different in other cultural or geographical areas. Thus, further work is needed to replicate the results in other geographical areas and other software development cultures. However, to mitigate this validity issue, we attempted to enlarge the respondents' sample by inviting additional participants globally via LinkedIn. Reliability addresses whether a study would yield the same results if other researchers replicated it. In this sense, triangulation is important [37], and to mitigate this threat, we have employed source triangulation (several companies and several professional roles), methodological triangulation (quantitative analysis based on surveys and qualitative analysis based on interviews), and observer triangulation (all authors participated in the analysis).

## 8. CONCLUSIONS

This study investigates how common different management strategies are when managing TD and investigating how software practitioners perceive such TD management strategies. More specifically, we study how software management influences how software practitioners work with TD, for example, by continuously encouraging and rewarding those who focus on TD remediation and limitation activities. Yet another TD management strategy we examine in this study is based on penalizations and forcing mechanisms. The results show that software practitioners are not commonly rewarded, penalized, or forced to keep the level of TD down.
Further, the result shows that a TD management strategy based on encouraging activities is described as having a significant impact on software engineers' attitudes and behaviors when addressing TD.

The results from both the first and the second step of this study show that an extensive part of the respondents state that they are not directly encouraged by managers to keep TD down. This indicates that there is considerable unfulfilled potential to influence how software practitioners can limit and reduce TD by adopting a TD management strategy based on encouraging activities where, for example, the concept of TD is acknowledged and recognized more broadly.
Moreover, this study also contributes to a TD management quadrant model describing four different TD management strategies and its tactics together with recommendations on how to implement such strategies in practice.
Finally, besides indicating the importance of that managers encourage the technical roles to address TD, it is also important that managers dedicate extra time and resources so that the technical roles actually may conduct these activities in reality.

# APPENDIX A

**Characteristics of the sample survey in Step 1**

| Factor | Percentage split | |
|---|---|---|
| **Experience** | < 2 years | 3,90% |
| | 2 - 5 year | 10,50% |
| | 5 - 10 year | 17,40% |
| | > 10 years | 68,20% |
| **Roles** | Developer/Program/Software Engineer | *49,20% |
| | Software Architect | 24,80% |
| | Manager | 6,20% |
| | Project Manager | 6,20% |
| | Product Manager | 5,0% |
| | Expert | 5,0% |
| | | 3,50% |
| **Team size** | 1–5 members | 23,30% |
| | 6–10 members | 36,00% |
| | 11–20 members | 15,90% |
| | 21–40 members | 6,60% |
| | > 40 members | 18,20% |

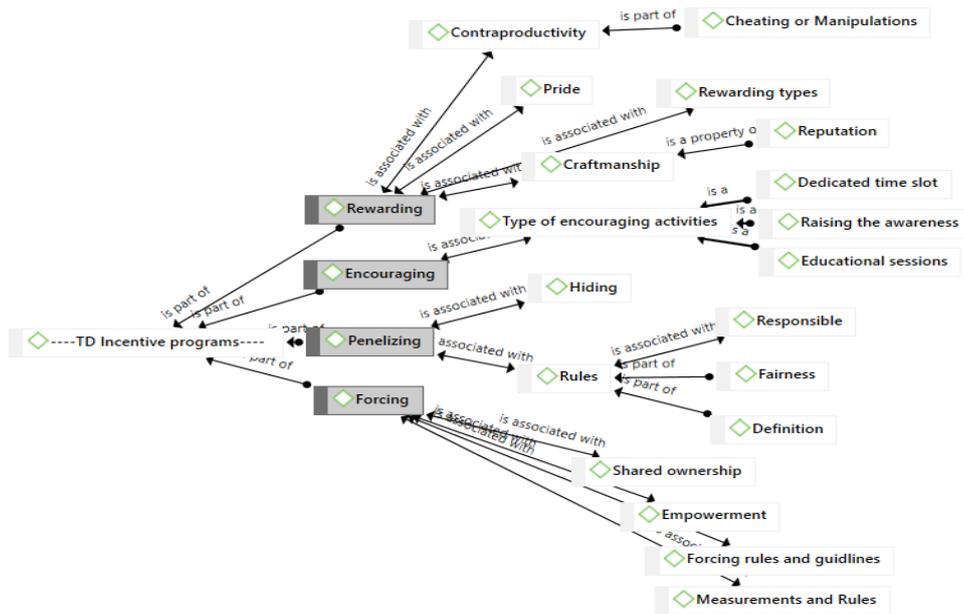

**Thematical Coding Scheme – Step 1**

## Examples of interview questions step 1

| TD Management Strategy | Examples of Interview Questions |
|---|---|
| **Encouragement (RQ1)** | • How do you perceive encouragement from managers for keeping the level of TD down?<br>• How could such a strategy be implemented?<br>• Do you agree with the results of the survey? |
| **Rewarding incentive (RQ2)** | • How do you perceive a rewarding incentive for keeping the level of TD down?<br>• Do you have this or a similar strategy in place or planned for the future?<br>• How could such a strategy be implemented? |
| **Forcing (RQ3)** | • How do you perceive a forcing mechanism for keeping the level of TD down?<br>• Do you have this or a similar strategy in place or planned for the future? |
| **Penalizing disincentive (RQ4)** | • How do you perceive a penalizing disincentive for keeping the level of TD down?<br>• How could such a strategy be implemented?<br>• Does your company have a team and/or personal incentive/disincentive system for any other kind of quality criteria related to your software development process? |
| **Other** | • Which strategy of keeping the TD down, do you consider to be the most/least successful (and why), and under what circumstances? |

## Characteristics of the sample survey step 2

|  | *Technical Roles* | | *Managers* | |
|---|---|---|---|---|
|  | Percentage split | Factor | Percentage split | Factor |
| **Experience** | < 2 years<br>2 - 5 year<br>5 - 10 year<br>10 – 20 years<br>< 20 Years | 4,3%<br>6,5%<br>4,3%<br>84,4<br>0 | < 2 years<br>2 - 5 year<br>5 - 10 year<br>10 – 20 years<br>< 20 Years | 11,5%<br>0%<br>7,7%<br>46,2%<br>34,6% |
| **Roles** | Developer<br>Team / FunctionalArchitect<br>Test / Quality<br>Platform / Chief Architect | 13%<br>23,9%<br>8,7%<br>54,3% | R&D manager<br>Product manager<br>CPO<br>Product Owner<br>Other managers | 38,5%<br>26,9%<br>15,4%<br>11,5%<br>7,7% |
| **Team size** | **Size of team:**<br>1–5 members<br>6–10 members<br>11–20 members<br>> 20 members | <br>23,9%<br>23,9%<br>52,2%<br>0 | **Managing numbers of teams:**<br>1–2 teams<br>3-5 teams<br>6–10 teams<br>11-15 teams<br>>15 | <br>53,8%<br>11,5%<br>7,7%<br>7,7%<br>19,2% |